\documentclass{ws-procs9x6}

\begin{document}

\title{Inclusive jet cross section measurement at D\O}

\author{M. VOUTILAINEN For the D\O\ collaboration\\
University of Nebraska in Lincoln, \\
Helsinki Institute of Physics}

\address{Fermilab --- D\O\ M.S. 357, \\
Batavia, IL 60510-0500, USA\\ 
E-mail: mavoutil@fnal.gov}

\maketitle

\abstracts{We present a new preliminary measurement of the inclusive jet cross section in $p\bar{p}$ collisions based on a integrated luminosity of about $0.8\,{\rm fb}^{-1}$. The data were acquired using the D\O\ detector between 2002 and 2005. Jets are reconstructed using an iterative cone algorithm with radius $R_{\rm cone} = 0.7$. The inclusive jet cross section is presented as a function of transverse jet momentum and rapidity. Predictions from perturbative QCD in next-to-leading order, plus threshold corrections in 2-loop accuracy describe the shape in the transverse jet momentum.}

\section{Introduction}

The production of particle jets in hadron collisions is described by the theory of Quantum Chromodynamics (QCD). When the transverse jet momentum with respect to the hadron beam direction ($p_T$) is large, the contributions from long-distance physics processes with low $p_T$ are small and the production rates of jets can be predicted by perturbative QCD (pQCD). The inclusive jet cross section in $p\bar{p}$ collisions at large $p_T$ is directly sensitive to the strong coupling constant ($\alpha_s$) and the parton density functions (PDFs) of the proton. Furthermore, potential deviations from the  pQCD prediction at high $p_T$, not explained by PDFs, may indicate new physics beyond the Standard Model.

\section{Jet energy scale}

The inclusive jet cross section is measured in two central rapidity regions $|y_{jet}| < 0.4$ and $0.4 < |y_{jet}| < 0.8$. We note that the data are corrected for underlying events when the jet energy scale is computed using
\begin{equation}
E_{ptcl} = \frac{E_{cal} - O}{R \cdot S},
\end{equation}
where $O$ is the offset contribution, $R$ is the response of the particle jet and $S$ is the net showering due to detector effects.
The jet energy scale corrects for ``offset" energies measured using zero-bias events which correspond to uranium noise (jets are measured with a liquid Argon-Uranium calorimeter), pile-up effects and underlying events because these effects cannot be distinguished experimentally. 
The electromagnetic calorimeter is calibrated using the Z peak in Z$\rightarrow e^{+}e^{-}$ events. The energy scale of the electromagnetic calorimeter is transferred to photons in $\gamma$+jets events, with photon purity and relative photon--electron energy scale mostly compensating each other. The absolute jet energy scale is assigned using the transverse momentum balance between the jet and the photon in the $\gamma$+jets events. The detector pseudorapidity dependence of the jet energy scale was determined using both dijet and isolated photon plus jet events.
Detector effects cause some of the particle jets energy to be showered outside the cone, or outside energy to be showered inside the cone. The net effect is accounted for by measuring the energy density profile around the jet, subtracting from this the contribution to the showering due to physics effects from Monte Carlo.

\section{Jet $p_{T}$ resolution and unfolding}

Jet $p_{T}$ resolution is measured on a subsample of the same dataset as is used for cross section measurement using dijet asymmetry
\begin{equation}
A = \frac{|p_{T,1}-p_{T,2}|}{p_{T,1}+p_{T,2}},
\end{equation}
which is corrected for soft radiation of additional jets below jet reconstruction threshold and for particle level imbalance.

Spectra in $p_T$  are fit, in an iterative procedure, with parameterized ansatz functions,
\begin{equation}
f(N,\alpha, \beta, \gamma) = N(p_{T}/\mathrm{GeV})^{-\alpha}\left(1 - \frac{2\cosh(y_{min})p_{T}}{\sqrt{s}}\right)^{\beta}\exp(-\gamma p_{T}),
\end{equation}
where $y_{min}$ is the minimum absolute rapidity in the bin and $\sqrt{s}$ is the center-of-mass energy, and folded with resolutions determined from data. Ratios of the original to the folded ansatz functions are used to correct the data for folding of resolution effects. The results were cross-checked with another method using PYTHIA\cite{pythia} smeared according to the jet $p_T$ and $y$ resolutions.

\section{Results}

\begin{figure}[ht]
\centerline{
  \includegraphics[width=0.5\textwidth]{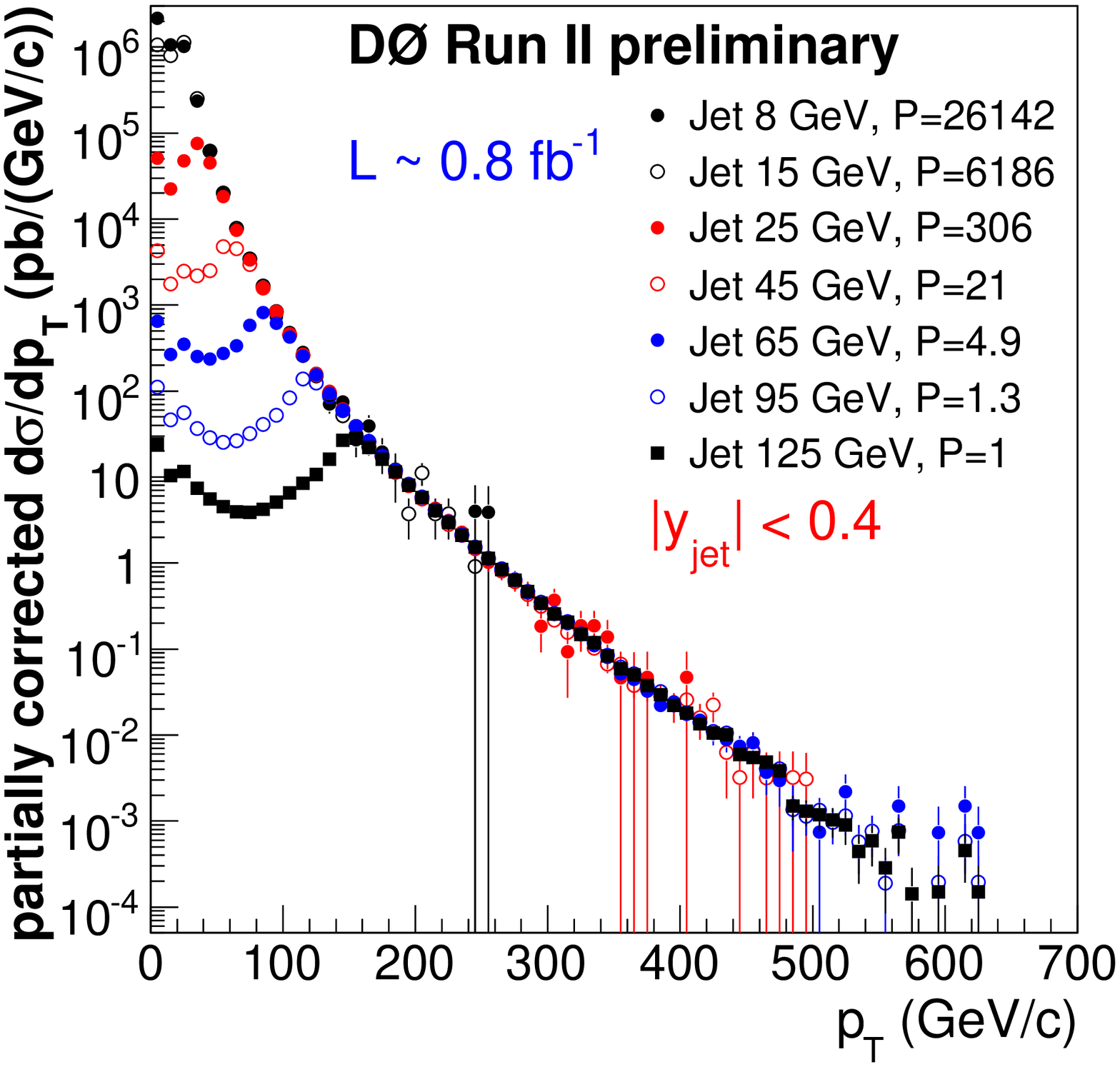}
  \includegraphics[width=0.5\textwidth]{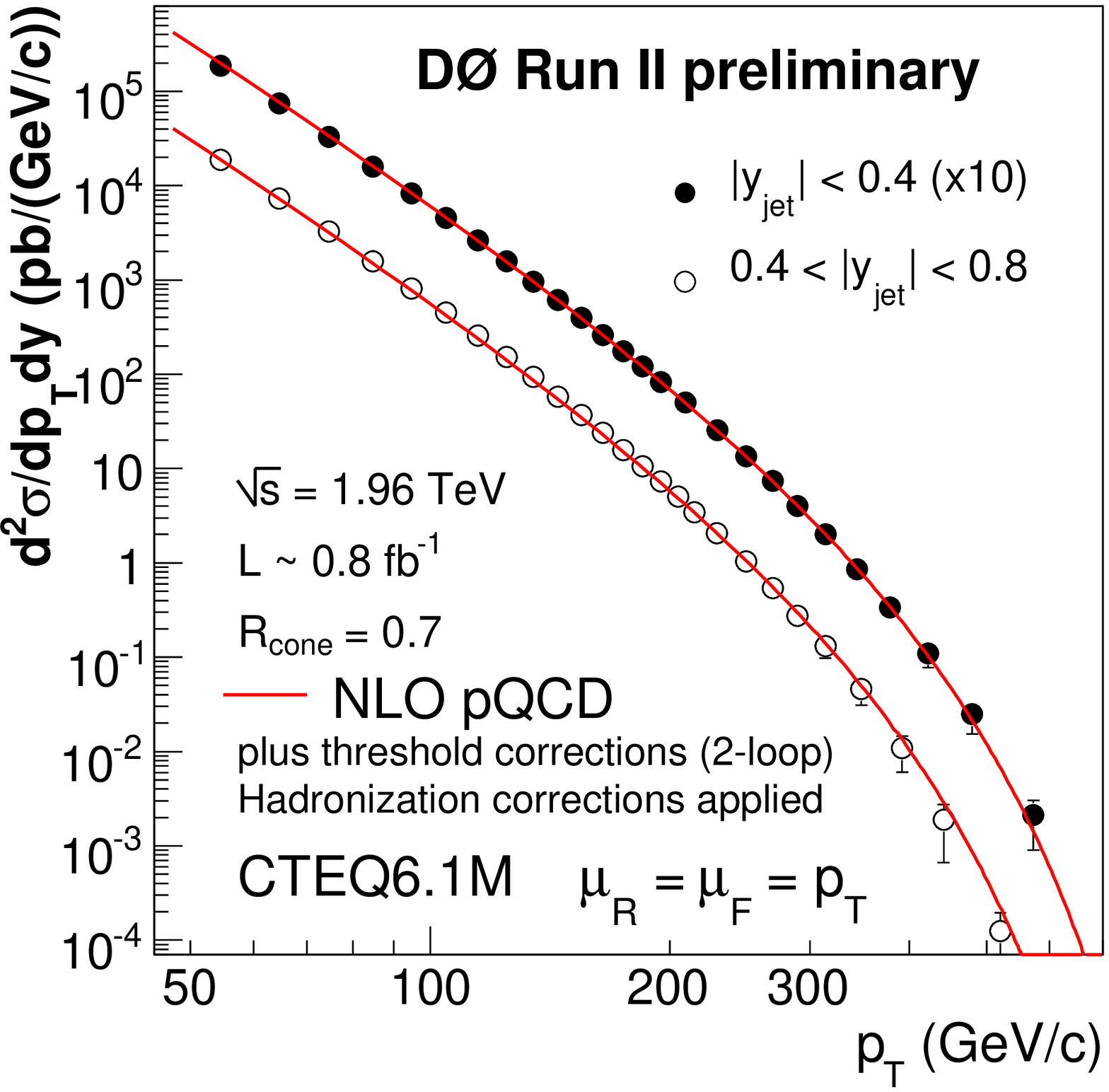}
}
\caption{Partially corrected inclusive jet cross section in central rapidity, measured with different jet triggers at different $p_{T}$ thresholds (left). The inclusive jet cross section, measured in two regions of jet rapidity (right). Error bars show the total measurement uncertainty. The predictions from pQCD are corrected for hadronization effects and are overlaid on the data as lines.\label{fig:triggers}\label{fig:final_xsec}}
\end{figure}

The cross section is measured using seven different jet triggers, as shown in Fig.~\ref{fig:triggers}. The different triggers are matched using relative trigger efficiencies and the cross section is corrected for jet identification and event selection efficiencies. The partially corrected spectrum and the final result corrected for $p_{T}$ resolution are shown in Fig.~\ref{fig:final_xsec}. The measurement is normalized to theory at $p_{T}$=100~GeV/c in $|y_{jet}|<0.4$ to remove luminosity uncertainty. The agreement with pQCD is good over the wide $p_{T}$ region explored.

\begin{figure}[ht]
\centerline{
  \includegraphics[width=0.5\textwidth]{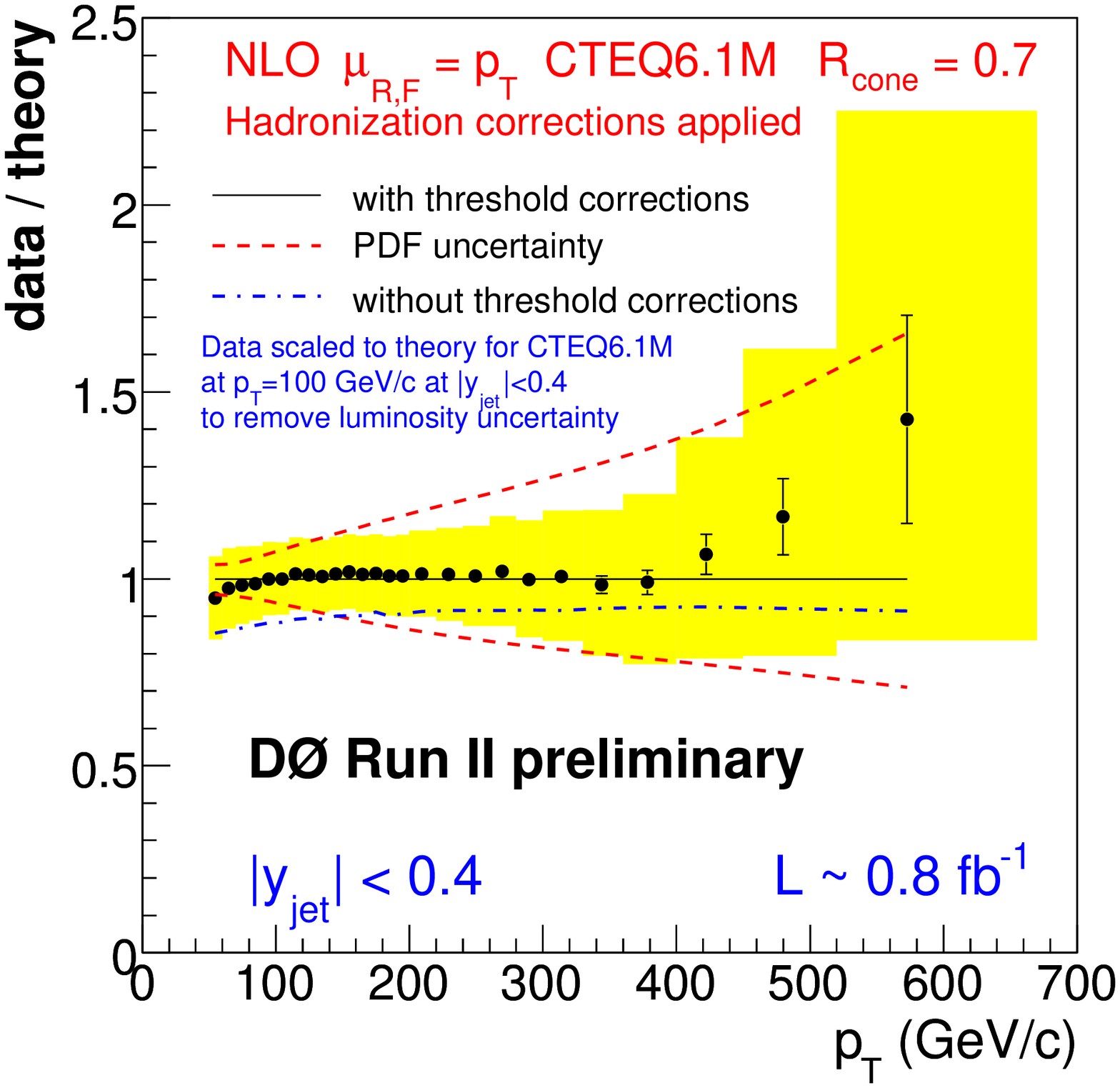}
  \includegraphics[width=0.5\textwidth]{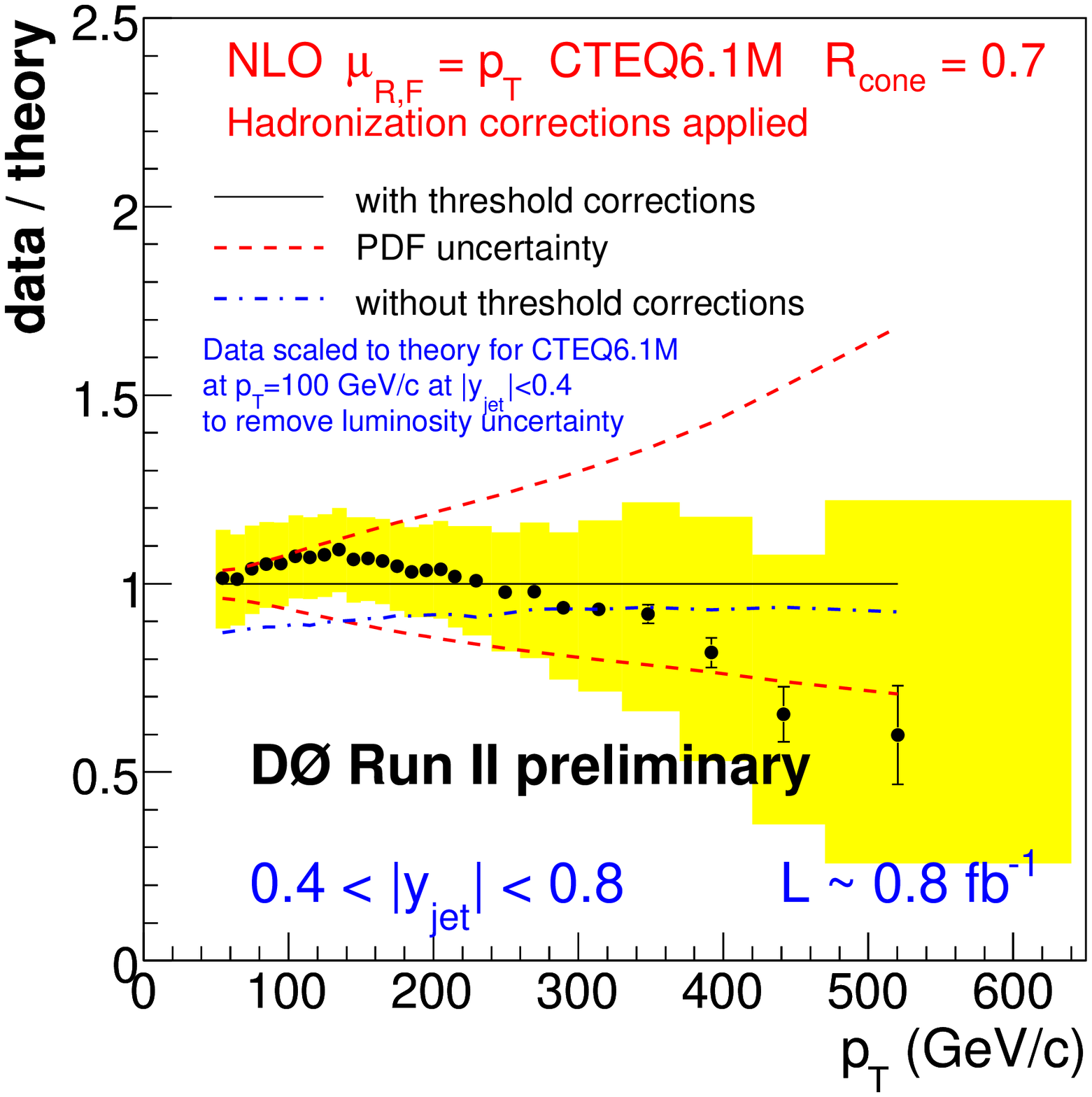}
}
\caption{Inclusive jet cross section over theory, measured in two regions of jet rapidity. Error bars and band show statistical and systematic uncertainty, respectively.\label{fig:comparison}}
\end{figure}

The cross section measurement is compared to next-to-leading order (NLO) theory with threshold corrections in 2-loop approximation\cite{kidonakis} in Fig.~\ref{fig:comparison} in two regions of jet rapidity. The NLO calculations were performed using NLOJET++\cite{nlojet} and fastNLO\cite{fastnlo}. The PDF uncertainty from CTEQ6.1M\cite{cteq6} is overlaid as dashed lines, showing that the measurement at high $p_{T}$ is getting precise enough to constrain the PDFs. Most of the PDF uncertainty at high $p_{T}$ is coming from the uncertainty in the gluon PDF at high momentum fraction $x$. The next-to-leading order theory without threshold corrections is also shown as a dash-dotted line.

\section{Conclusion}

Preliminary results on the inclusive jet cross section measurements at D\O\ were shown with 0.8~fb$^{-1}$ of collected luminosity. The results are in good agreement with the next-to-leading order perturbative QCD calculations. The measurement is sensitive to the quark and gluon content in the proton and allows one to reduce the gluon density uncertainty at high momentum fraction. This is one of the leading limitations of beyond Standard Model searches at the Tevatron and the LHC.

\section*{Acknowledgements}

I thank my colleagues at D\O\ and acknowledge support from the Graduate School in Particle and Nuclear Physics, the Finnish Cultural Foundation and Magnus Ehrnrooth Foundation.

\end{document}